\documentclass[twocolumn,showpacs,preprintnumbers,amsmath,amssymb,aps]{revtex4-1}

\usepackage{cases}
\usepackage{bm}
\usepackage{graphicx}
\usepackage{epsfig}
\usepackage{hyperref}
\usepackage{xcolor}
\usepackage[english]{babel}

\begin{document}

\title{Flat band superconductivity in the square-octagon lattice}

\author{Lizardo H. C. M. Nunes}
\email{LizardoNunes@ufsj.edu.br}
\altaffiliation{%
Departamento de Ci\^encias Naturais,
Universidade Federal de S\~ao Jo\~ao del Rei, 36301-000
S\~ao Jo\~ao del Rei, MG, Brazil
}
\author{Cristiane Morais Smith}
\email{C.deMoraisSmith@uu.nl}
\affiliation{%
Institute for Theoretical Physics,
Center for Extreme Matter and Emergent Phenomena,
Utrecht University,
Princetonplein 5, 3584 CC Utrecht,
the Netherlands
}

\date{\today}

\begin{abstract}
  The discovery of superconductivity
  in twisted bilayer graphene
  has triggered a resurgence of interest
  in flat-band superconductivity.
  Here, we investigate the square-octagon lattice,
  which also exhibits two perfectly flat bands
  when next-nearest neighbour hopping
  or an external magnetic field
  are added to the system.
  We calculate the superconducting phase diagram
  in the presence of on-site attractive interactions
  and find two superconducting domes,
  as observed in several types
  of unconventional superconductors.
  %
  %
  %
  %
  The critical temperature
  shows a linear dependence
  on the coupling constant,
  suggesting that superconductivity
  might reach high temperatures
  in the square-octagon lattice.
  Our model could be experimentally realized
  using photonic or ultracold atoms lattices.
\end{abstract}


\maketitle 

\section{Introduction}
\label{Sec.Introduction}

It has been conjectured
that the presence of flat bands
in a two-dimensional system may give rise
to room temperature superconductivity~\cite{Volovik2018}.
%
Indeed,
while for a conventional BCS superconductor,
the critical temperature scales exponentially
with the inverse of the interaction strength,
for a flat band system
the critical temperature
exhibits a linear dependence
on the interaction~\cite{Heikkla2011,Kopnin2011},
indicating a robust superconducting phase.
The BCS result
for narrow bands
in the strong coupling limit
becomes
$ T_c \propto g N( 0 ) $
where $ N(0) $
is the density of the states
at the Fermi level,
which is enhanced
as the flat band
leads to maximal values
of the density of states~\cite{Marchenko2018}.

The conjecture seems to be ratified
by the discovery of superconductivity
in twisted bilayer graphene~\cite{Cao2018a}:
when two stacked sheets of graphene
are twisted relative to each other
by about 1.1 degrees,
the so-called first ``magic" angle,
zero-resistance states
with critical temperature of up to 1.7 K
arise upon electrostatic doping.
This emergent superconductivity is absent
in a single layer graphene
and occurs because the twisting
leads to the formation
of a Moir\'e pattern,
and a consequent shift
of the Van Hove singularity
to the Fermi energy.
This phenomenon has been predicted theoretically
a few years ago~\cite{Bistritzer2011},
but has been experimentally observed
only recently~\cite{Cao2018a,Cao2018b}.
The general understanding is that
due to the presence
of flat bands,
the kinetic energy is quenched
and  interaction-driven quantum phases prevail.
A similar explanation
has been proposed
to interpret the appearance
of high-$ T_c $ superconductivity
in highly oriented pyrolytic graphite~\cite{Esquinazi2014,Kopelevich2015}.

The possibility
to access flat bands
and their influence
on the physical properties
of the system
have been studied
for about three decades~\cite{Lieb1989,Mielke1991,Tasaki1992,Arita2002,Tanaka2003,Noda2009,Katsura2010,Julku2016,Hartman2016,Kumar2017}
and recently
there is a resurgence of interest
in flat bands
to explore unconventional superconductivity~\cite{Peotta2015,Liang2017,Aoki2019,Kumar2019,Hoffmann2019,Sayyad2020}.
In fact, given
the implementation of experiments
with cold atoms, photons or electrons
in the micro and nanoscale respectively,
many of the long-standing theoretical predictions
for the flat band systems
may finally be tested experimentally.
Indeed,
flat bands have been observed
not only in electronic systems
or spin chains with frustration~\cite{Derzhko2015,Chalker2011},
but also in artificial lattices~\cite{Leykam2018a},
as in ultracold atomic gases~\cite{Taie2015,Ozawa2017,Taie2018,An2018}
or photonic devices~\cite{Schulz2017,Maczewsky2017,Mukherjee2017,Real2017,Klembt2017,Whittaker2018,Leykam2018b}.

Here
we investigate the superconducting phase
of the square-octagon lattice,
which is the arrangement of the octagraphene~\cite{Sheng2012}.
The square-octagon lattice
has been attracting much attention lately
due to the plethora of novel phases
predicted to occur in the system.
They range from
a quantum magnetic phase
under the competition between temperature
and on-site repulsive interaction~\cite{Bao2014},
to topological insulating phases
induced by spin-orbit coupling
or non-Abelian gauge fields~\cite{Yang2019,Yang2018,Kargarian2010},
and even high-temperature superconductivity
with singlet s$_{ \pm } $-wave paring symmetry~\cite{Kang2019}.
Moreover,
it has been shown that
trivial and nontrivial flat bands
can be tuned
on the square octagon lattice
by considering the addition
of next-nearest neighbour hopping
and an external magnetic flux~\cite{Biplab2018,Sil2019}.

In this paper,
we calculate the superconducting phase
due to on-site attractive interactions
when the system presents perfectly flat bands.
We obtain
multiple superconducting domes,
as observed
in several strongly correlated compounds~\cite{Das2016}
and twisted graphene bilayer~\cite{MacDonald2019}.
Since we analyse the conditions
for the appearance of superconductivity in the system,
%
%
our conclusions
may shed some light
in the understanding of the experimental data
for those unconventional superconductors as well.


This paper is structured as follows:
in Sec.~\ref{Sec.H0}
we analyse the conditions
for the appearance of flat bands
in the square-octagon lattice
assuming nearest neighbour (NN)
and next-nearest neighbour (NNN) hoppings
and the presence of an external magnetic flux.
Then, in Sec.~\ref{Sec.Onsite}
we calculate the superconducting phase diagram.
Our conclusions are presented
in Sec.~\ref{Sec.Conclusions}.

\section{Flat bands}
\label{Sec.H0}

\begin{figure}[t!]
  \centering
  \includegraphics[width=1\columnwidth]
  {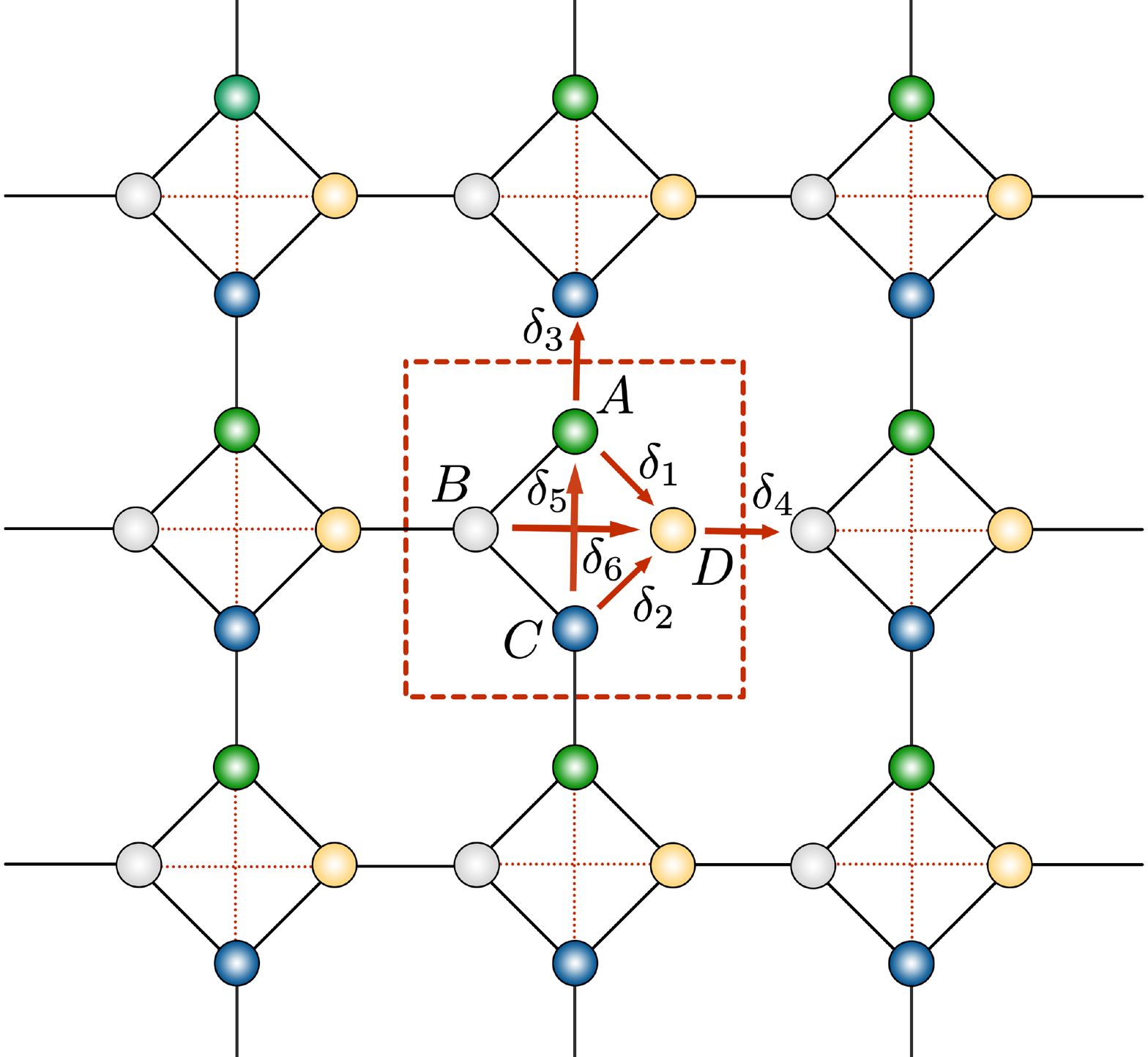}
  \caption{(Color online)
  The geometry of square-octagon lattice.
  $ A, B, C, D $ denote four different sites in the unit cell,
  which indicate a large square (red dashed lines).
  The strength of the NN hopping along the sides (black solid lines)
  of the small square is $ t $.
  The strength of the NNN hopping
  (red dotted lines)
  along the diagonals
  of the small square is $ \tau $.
  $ \delta_1, \delta_2, \delta_3, \delta_4 $
  are NN vectors
  and
  $ \delta_5, \delta_6 $
  are NNN vectors.
  }
  \label{Fig.SquareOctagenLattice}
\end{figure}

The tight-binding model
that describes the kinetic energy
of spin 1/2 fermions
in the square-octagon lattice
(see Fig.~\ref{Fig.SquareOctagenLattice})
can be written as~\cite{Sil2019},
\begin{equation}
H_0
=
\sum_{ m, n }
\sum_{ i, j }
\tau_{ i, j }
c^{ \dagger }_{ m, n, i, \sigma  } c_{ m, n, j, \sigma  }
+
\mbox{ h.c.}
\, ,
\label{Eq.H0}
\end{equation}
where
the operator
$ c^{ \dagger }_{ m, n, i, \sigma  } $
 ($ c_{ m, n, i, \sigma  } $)
creates (annihilates)
a fermion with spin
$ \sigma = \uparrow, \downarrow $
at the $ i $-th site of the $ (m, n) $-th unit cell.
$ \tau_{ i, j } $ is the hopping parameter
between the $ i $-th site and the $ j $-th sites,
and it can take two possible values,
depending on the position
of the sites $ i $ and $ j $.
For an electron hopping
along the boundary of a square plaquette,
$ \tau_{ i, j } = t $,
which is the NN hopping.
Along the lines
inside the square plaquettes,
$ \tau_{ i, j } = \tau $,
which denotes the NNN hopping.
As shown in Fig.~\ref{Fig.SquareOctagenLattice},
the NN vectors are defined as
$ {\bf \delta}_1 = \left( 1/ \sqrt{ 2 } \right) ( 1, -1 ) $,
$ {\bf \delta}_2 = \left( 1/ \sqrt{ 2 } \right) ( 1, 1 ) $,
$ {\bf \delta}_3 = ( 0, 1 ) $,
and
$ {\bf \delta}_4 = ( 1, 0 ) $,
while the NNN vectors are
$ {\bf \delta}_5 =  \sqrt{ 2 } ( 0, 1 ) $,
and
$ {\bf \delta}_6 = \sqrt{ 2 } ( 1, 0 ) $.
The lattice parameter
is set to unit for the sake of simplicity.
The lattice translation vectors are
$ {\bf a }_1 = ( 1 + \sqrt{ 2 }, 0 ) $
and
$ {\bf a }_2 = ( 0, 1 + \sqrt{ 2 } ) $,
so that each unit cell is given by
$ { \bf R  }(m,n) = m {\bf a }_1 + n {\bf a }_2  $ .

Now, we assume that
for each square plaquette
there are Aharonov-Bohn phase factors
incorporated to the hopping terms,
$ t \rightarrow t \, \exp(\, \pm i \Theta \,) $,
due to the presence of an external magnetic flux $ \Phi $
(or some artificial gauge field).
Here,
$ \Theta = \pi \, \Phi / 2 \Phi_0 $,
with $ \Phi_0 = h c / e $
denoting the fundamental flux quantum.
The positive and negative signs in the exponent
indicate the direction of the forward and backward hoppings,
respectively.
By performing
a Fourier transformation and
introducing the spinor
$
\psi_{ {\bf k} , \sigma }
=
  (
  c^{ A }_{ {\bf k} , \sigma },
  c^{ B }_{ {\bf k} , \sigma },
  c^{ C }_{ {\bf k} , \sigma },
  c^{ D }_{ {\bf k} , \sigma }
  )^{ T }
  $,
where
$ c^{ \alpha }_{ {\bf k} , \sigma } $
are the fermion annihilation operators
in the four basis of the unit cell
($ \alpha = A, B, C, D $, as in Fig.~\ref{Fig.SquareOctagenLattice}),
with
$ k_1 = { \bf k } \cdot { \bf a }_1 $
and
$ k_2 = { \bf k } \cdot { \bf a }_2 $,
the Hamiltonian in Eq.~(\ref{Eq.H0})
becomes
$ H
=
-
\sum_{ {\bf k }, \sigma }
\psi^{ \dagger }_{ {\bf k} , \sigma }
\,
\tilde{ H }_0
\,
\psi_{ {\bf k} , \sigma }
$,
where \cite{Biplab2018}
\begin{eqnarray}
\tilde{ H }_0
=
\left(
\begin{array}{cccc}
0 & t e^{ i \Theta } & t e^{ i k_2 } + \tau  &  t e^{ - i \Theta }
\\
t e^{ - i \Theta } & 0 & t e^{ i \Theta } &  t e^{ - i k_1 } + \tau
\\
t e^{ - i k_2 } + \tau & t e^{ -i \Theta }  & 0 & t e^{ i \Theta }
\\
t e^{ i \Theta } &  t e^{ i k_1 } + \tau & t e^{ -i \Theta } & 0
\\
\end{array}
\right)
\, .
\label{Eq.H0.tilde}
\end{eqnarray}
From now on,
we set $ t = 1 $
throughout this work
without any loss of generality,
and $ \tau $ is given in units of $ t $.

The particular case
$ \Theta = \tau = 0 $,
i.e., when
there is no magnetic flux
and NNN hopping is neglected,
has been previously investigated
by Yamashita {\it et al.}\cite{Yamashita2013}.
It was shown
that the system is metallic
and there are Dirac cones
in the first Brillouin zone,
with flat bands only
along the lines
$ k_1 = \pi $
or
$ k_2 = \pi $.

The band solutions $ \epsilon $
for the generic case of finite
$ \tau $ and $ \Theta \neq 0 $
can be quite complicated,
since they are provided
by the solutions of the equation
\begin{equation}
 |\, \tilde{ H }_0 - \epsilon {\bf 1}  \, |
=
  F
  +
  G
  =
  0
  \label{Eq.Dispersion}
\end{equation}
for each spin channel,
where
\begin{eqnarray}
F
& = &
\epsilon^4
-
2 \epsilon^2
    \left[\,
            \tau^2 + \tau \, f( k_1, k_2 ) + 3
    \,\right]
\nonumber
\\
& & -
  4 \epsilon \cos 2\Theta
  \left[\,
          f( k_1, k_2 ) + 2 \tau
  \,\right]
\, ,
\label{Eq.F}
\end{eqnarray}
\begin{eqnarray}
G
& = &
\tau^4
+ 2 \tau^3 f( k_1, k_2 )
+
2 \tau^2 \left(\,  2 \cos k_1 \cos k_2 - 1 \,\right)
\nonumber
\\
&  & -
2 \tau f( k_1, k_2 ) - 2 \cos 4 \Theta + 3
\nonumber
\\
& &
-
4 \cos k_1 \cos k_2
\, ,
\label{Eq.G}
\end{eqnarray}
and $ f( k_1, k_2 ) = \cos k_1 + \cos k_2 $.

However,
a simple possible condition
for the appearance of flat bands in the model
is obtained for $ G = 0 $.
In this case,
one can see
from Eq.~(\ref{Eq.F})
that there is
at least one flat band,
at $ \epsilon = 0 $.
Let us consider for the moment
$ \Theta = 0 $.
Under this constraint,
there are four values
for the hopping $ \tau $
satisfying $ G = 0 $:
$ \tau = \pm 1, \tau_{ \pm } $,
where
\begin{equation}
  \tau_{ \pm }
  =
  f( k_1, k_2 )
  \pm
  \sqrt{
    \frac{ f( 2 k_1, 2 k_2 ) - 4( \cos k_1 \cos k_2 -1 )
    }{
     2 }
     }
  \, .
  \label{Eq.Tau}
\end{equation}
The roots $ \tau_\pm $
indeed correspond to a flat band $ \epsilon = 0 $,
but since the solutions
are functions of $ k_1 $ and $ k_2 $,
the adjustable parameter
$ \tau = \tau_\pm $
cannot be easily used in experiments
in order to engineer
a model with flat bands.
%
In contrast,
for $ \tau = \pm 1 $,
we obtain not just one,
but rather two flat band solutions
from Eq.~(\ref{Eq.G}):
For $ \tau = - 1 $,
there are flat bands at
\begin{equation}
\epsilon_{1, 2 } =  2, 0 \, ,
\end{equation}
and dispersive bands at
\begin{equation}
\epsilon_{ 3, 4 } =  -1 \pm \sqrt{ 5 - 2 f( k_1, k_2 ) }
\, ,
\end{equation}
whereas for
$ \tau = 1 $
the bands are
\begin{equation}
\epsilon
=
    \begin{cases}
      \,  0  \, , \\
      \, - 2  \, ,\\
      \, 1 \pm \sqrt{ 5 + 2 f( k_1, k_2 ) }.
    \end{cases}
    \label{Eq.DispersionResults.Tau1.Theta0}
\end{equation}

\begin{figure}[t!]
  \centering
  \includegraphics[width=1\columnwidth]
  {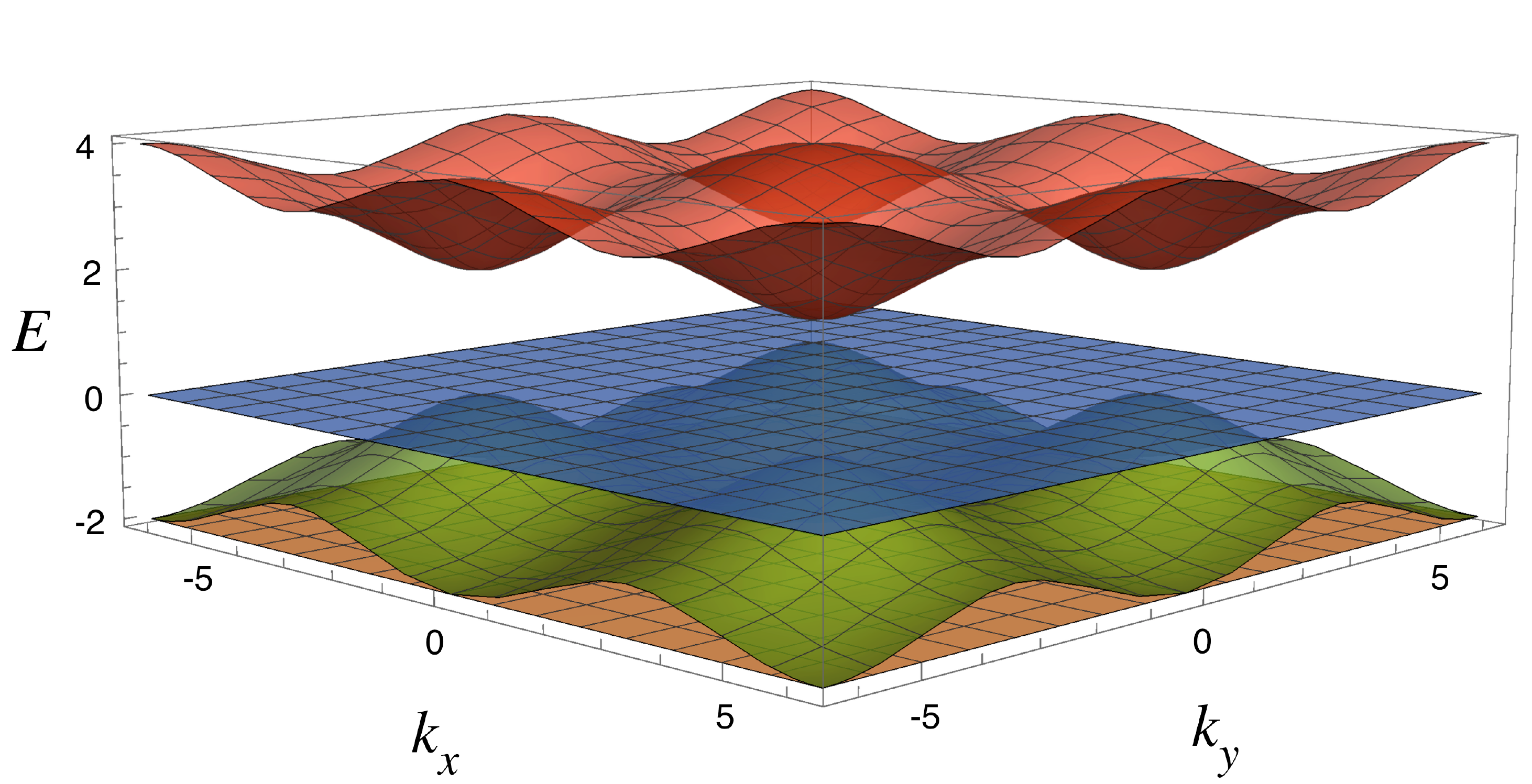}
  \caption{(Color online)
   Band structure
  of the noninteracting tight-binding
  square-octagon lattice model
  in momentum space.
  $ \tau =1 $,
  $ \Theta = 0 $,
  and the energies are expressed in units of $ t $.
  }
  \label{Fig.H0Bands}
\end{figure}

The band structure
for $ \tau =  1 $ and $ \Theta = 0 $
has been previously investigated~\cite{Biplab2018,Sil2019}
and a remarkable feature for the square-octagon lattice
is that one of the dispersive bands
is sandwiched in between
two perfectly flat bands,
while the other is isolated from the rest,
at the top of the spectrum,
as depicted in Fig.~\ref{Fig.H0Bands}.

As emphasized in Ref.~\cite{Biplab2018},
this is a remarkable result,
very different from the usual flat bands
that appear only
at the maximum or minimum of the spectrum
in absence of any magnetic field~\cite{Ohgushi2000,Bergman2008}.

Notice that,
within the first Brillouin zone,
the sandwiched dispersive band
touches the higher flat band
($ \epsilon = 0 $)
at $ {\bf k} = ( \pm \pi, \pm \pi ) $
and the lowest flat band
at $ {\bf k} = ( 0, 0 ) $.

Moreover,
numerical calculations~\cite{Biplab2018}
indicate that
flat bands appear
even in the presence of a magnetic flux
and remain robust
for fairly high values of the flux,
although
they
are not observed for every non-zero value of $ \Theta $.
Indeed,
our studies ratify
the numerical results,
since it can be
promptly verified analytically
from Eq.~(\ref{Eq.Dispersion})
that perfectly flat bands arise
for $ \tau = 1 $ and $ \Theta \neq 0 $,
\begin{equation}
 |\, \tilde{ H }_0 - \epsilon {\bf 1}  \, |
=
F - 2 \left(\, \cos 4 \Theta - 1 \,\right)
= 0
\, .
  \label{Eq.Dispersion.Tau1.ThetaNeq0}
\end{equation}
Thus,
whenever $ \cos 4 \Theta = 1$,
one obtains at least one flat bland,
$ \epsilon = 0 $,
and hence
$ \Theta = n \, \pi / 2 $,
with
$ n \in \mathbb{ Z } $.
In particular,
for this specific choice of
$ \Theta $ and $ \tau = 1 $,
we get two flat bands
and also two other dispersive bands,
\begin{equation}
\epsilon
=
    \begin{cases}
      \,  0   \, ,\\
      \, 2 l  \, ,\\
      \, - l \pm \sqrt{ 5 + 2 f( k_1, k_2 ) } \, ,
    \end{cases}
    \label{Eq.DispersionResults.Tau1.ThetaNeq0}
\end{equation}
where $ l = \pm 1 $
for odd or even values of $ n $, respectively.
Therefore,
from now on,
we set
$ \tau = 1 $
and
$ \Theta = n \pi /2 $
for even values of $ n $
in the remaining of this paper.

Although
we have concentrated here
on the case $ \tau = 1 $ and $ n $ even
for the magnetic flux quantization,
$ \Theta = n \pi /2 $,
the results for $ n $ odd
or $ \tau = -1 $
are essentially the same,
since one still obtains
two perfectly flat bands
and two dispersive bands,
where one of the bands is detached
from the others
and the remaining
is sandwiched between the two flat bands.

Finally,
the Chern numbers
have been calculated
for each of the bands~\cite{Biplab2018}
with $ \Phi \neq 0 $
and, as expected,
the dispersive bands are trivial,
but nearly flat bands
present nonzero Chern numbers.

\section{On-site pairing interactions}
\label{Sec.Onsite}

Let us now turn our attention
to the effect of interactions in the system
by introducing an on-site
Hubbard-like attractive term,
expressed as
$ - g \, n^{\alpha }_{ \uparrow } n^{\alpha }_{ \downarrow } $,
where
$ g > 0 $
is the interaction strength
and
$ n^{\alpha }_{ \sigma } $
is the number operator for fermions,
with $ \alpha = A, B, C, D $
labelling the basis of the unit cell,
as before.
Taking into account every site of the lattice,
the interaction Hamiltonian
can be rewritten as
\begin{equation}
H^{ \mbox{\scriptsize{on-site}} }_{ int }
=
- g
\sum_{ \alpha, i }
\left( p^{ \alpha }_i \right)^{ \dagger } p^{ \alpha }_i
\, ,
\label{Eq.H.int.onSite}
\end{equation}
where the operator
$
p^{ \alpha }_i
=
c^{ \alpha }_{ i, \downarrow } c^{ \alpha }_{ i, \uparrow }
$
destroys a Cooper pair at the $ i $-th site of the lattice
and its adjoint
$ \left( p^{ \alpha }_i \right)^{ \dagger } $
creates a Cooper pair.

In a flat band,
the interactions dominate
over the kinetic energy,
which suggests that
a mean-field calculation
might be appropriate
to describe the superconducting phase.
Therefore,
the Fourier transform
of Eq.~(\ref{Eq.H.int.onSite})
becomes
\begin{equation}
\tilde{ H }^{ \mbox{\scriptsize{on-site}} }_{ int }
=
-
\sum_{ \alpha, \bf k }
\left[\,
\left( p^{ \alpha }_{\bf k} \right)^{ \dagger } \Delta
+
p^{ \alpha }_{\bf k} \Delta^{ * }
-
\frac{  | \Delta |^2 }{ g }
\, \right]
\, ,
\label{Eq.H.int.onSite.MF}
\end{equation}
where we have introduced
the superconducting order parameter,
$
\Delta
= - g \sum_{ \bf k }
c^{ \alpha }_{ -\bf k, \downarrow }c^{ \alpha }_{  \bf k, \uparrow }
$
and we have also assumed that
$ \Delta^{ \alpha } \equiv \Delta $.

Now,
taking the spinor
$ \psi_{ \bf k, \sigma  } $
introduced above
and defining the Nambu operator
$
\Psi^{ \dagger }_{ \bf k } =
\left(\,
\psi^{ \dagger }_{ \bf k, \uparrow  }
; \,
\psi^{ T }_{ \bf k, \downarrow  }
\,\right)
$,
one can combine
the noninteracting Hamiltonian
in Eq.~(\ref{Eq.H0.tilde})
with the interaction term
in Eq.~(\ref{Eq.H.int.onSite.MF}),
so that
our model Hamiltonian
becomes
$
H
=
\sum_{ {\bf k } }
\Psi^{ \dagger }_{ {\bf k}  }
\,
\tilde{ H }
\,
\Psi_{ {\bf k} }
$,
where
\begin{eqnarray}
\tilde{ H }
& = &
\left(
\begin{array}{cc}
\tilde{ H }_{ 0 } - \mu & H_{ \Delta }
\\
H_{ \Delta^{ * } } &  - \tilde{ H }_{ 0 } + \mu
\end{array}
\right)
\, ,
\label{Eq.H.tilde}
\end{eqnarray}
and we have introduced the chemical potential $ \mu $,
with
$ H_{ \Delta } = \Delta \times {\bf 1}_4 $,
where
$ {\bf 1}_4  $
is the 4 $ \times $ 4 identity matrix.
Therefore,
the eigenvalues of our model Hamiltonian are
\begin{subnumcases}
  {
  E =
  \label{Eq.E}
  }
  \pm \sqrt{\, \mu^2 + | \Delta |^2 \,}
  \, ,
  \label{Eq.Eflat0}
  \\
  \pm \sqrt{ \left( 2 + \mu \right)^2 + | \Delta |^2 }
  \, ,
  \label{Eq.Eflat2}
  \\
  \pm \sqrt{ \left(\, | 1 - \mu | - \xi \,\right)^2 + | \Delta |^2 }
  \, ,
  \label{Eq.EdispSand}
  \\
  \pm \sqrt{ \left(\, | 1 - \mu | + \xi \,\right)^2 + | \Delta |^2 }
  \, ,
  \label{Eq.EdispUpper}
\end{subnumcases}
where
$ \xi = \sqrt{ 5 + 2 f( k_1, k_2 ) }$.
Notice that
for $ \Delta = \mu = 0 $
we get
$ E = \pm | \epsilon | $,
where  $ \epsilon $
are the energies
of the noninteracting system,
with $ l = -1 $
in Eq.(\ref{Eq.DispersionResults.Tau1.ThetaNeq0}).
In addition,
Eqs.~(\ref{Eq.Eflat0}) and (\ref{Eq.Eflat2})
is related to the flat bands obtained previously,
while
Eqs.~(\ref{Eq.EdispSand}) and (\ref{Eq.EdispUpper})
to the remaining dispersive bands.
Since the band given by Eq.~(\ref{Eq.EdispUpper})
is detached from the others,
we will consider
an effective model from now on
that disregards it.

The condition
for the appearance
of superconductivity
is provided by
the nonzero values of $ \Delta $
that minimize the free energy (effective potential),
given by
\begin{equation}
V_{ \rm{ eff } }
=
\frac{ | \Delta |^2 }{ g }
-
T \sum_n \,
\int
\frac{ d^2 k }{ \mathcal{ A } }
\,
\log\left[\,
\prod_{ j =  1}^{ 8 }
\left( i \omega_n - E_j \right)
\, \right]
\, ,
\label{Eq.Veff}
\end{equation}
where
$ T $ is the temperature,
$ \mathcal{ A } $ is
the area in momentum space,
$ \omega_n $ are the well known fermionic Matsubara frequencies
and $ j $ is the index
for the eigenvalues
in Eq.~(\ref{Eq.E}).
We have set
$ k_B = \hbar = 1 $
for the sake of simplicity.

Deriving
$ V_{ \rm{ eff } } $
with respect to $ \Delta $,
and performing the sum over
$ \omega_n $,
we arrive at the gap equation,
\begin{equation}
  1
  =
  \frac{ g }{2 }
  \sum_{ j {\text{ (odd ) }} }
  \int
  \frac{ d^2 k }{ \mathcal{ A } }
  \,
  \frac{ 1 }{ E_j }
  \tanh\left( \frac{ E_j }{ 2 T } \right)
  \, ,
  \label{Eq.GapEquation}
\end{equation}
where $ j {\text{ (odd ) }} $
indicates that
we are summing only over
the first three positive values of $ E $
in Eqs.~(\ref{Eq.Eflat0})-(\ref{Eq.EdispSand}).

From Eq.~(\ref{Eq.GapEquation}),
we obtain the self-consistent equation
for the critical temperature $ T_c $,
which is defined
as the temperature at which $ \Delta = 0 $,
\begin{equation}
  \frac{ 1 }{   \Delta_{ 0 0 } }
  =
  \frac{ 1 }{ | \mu | }
  \tanh\left( \frac{ | \mu | }{ 2 T_c  } \right)
  +
  \frac{ 1 }{ | 2 + \mu | }
  \tanh\left( \frac{ | 2 + \mu | }{ 2 T_c  } \right)
  +
 I\left( k_c \right)
\, ,
\label{Eq.Tc}
\end{equation}
where
\begin{equation}
  \Delta_{ 0 0 }
  \equiv
  \frac{ g }{ 2 }
  \left[\,
  \frac{ \mathcal{ A } (k_c) }{ \mathcal{ A } }
  \,\right]
  =
  \frac{ g }{ 2 }
  \left[\,
  \frac{ 1 }{ \mathcal{ A } }
  \int_{ | k | < k_c } d^2k
  \,\right]
  \, ,
  \label{Eq.Delta00}
\end{equation}
and
\begin{equation}
I\left( k_c  \right)
=
\frac{ 1 }{ \mathcal{ A } (k_c) }
\int_{ | k | < k_c } d^2k
\,
\frac{ 1 }{  | \xi_{ \mu } | }
\tanh\left( \frac{ | \xi_{ \mu } | }{ 2 T_c  } \right)
\, ,
\label{Eq.I}
\end{equation}
with $ \xi_{ \mu } =  \xi - | 1 - \mu | $,
and we have introduced
the momentum cutoff $ k_c $.

In order to calculate
$ T_c $
from Eq.~(\ref{Eq.Tc}),
three parameters have to be given:
$ \Delta_{ 00 } $, $ \mu $
and $ k_c $.
Notice that a natural cutoff
emerges from the theory,
depending on the microscopic mechanism
that is responsible for the Cooper pair formation.
In the BCS theory,
a natural cutoff
is the Debye frequency,
since the pairing is due
to the electron-phonon interaction.
Similarly,
for the square-octagon lattice,
a natural cutoff
is given
in terms of the inverse
of the lattice constant,
and we integrate the momentum
over the first Brillouin zone.
In such case,
$ \Delta_{ 00 } = g / 2 $.
Let us now expand
$
\xi
$
in polar coordinates for small values of $ k $,
$
\xi
\sim  \sqrt{  9 - k^2 }
$.
Introducing the
change of variables
$ x = ( k / k_c )^2 $,
we get
$
\xi( x )
\sim  3 \sqrt{ 1 - \delta_c x  }
$,
where the new parameter
$ \delta_ c = ( k_c / 3 )^2 $
incorporates the momentum cutoff $ k_c $.
From that,
one can
rewrite Eq.~(\ref{Eq.I}) as
\begin{equation}
I( \mu, T_c  )
=
\int_{ 0 }^{ 1 } dx
\,
\frac{ 1 }{  |\, \xi_{ \mu }( x ) \,| }
\tanh\left( \frac{ |\, \xi_{ \mu }( x ) \,| }{ 2 T_c  } \right)
\, .
\label{Eq.Ib}
\end{equation}
From now on,
we set $ \delta_c = 1 $,
which is equivalent to integrate
over the first Brillouin zone.


\subsection{Square-octagon lattice}

\begin{figure}[t!]
  \centering
  \includegraphics[width=1\columnwidth]
  {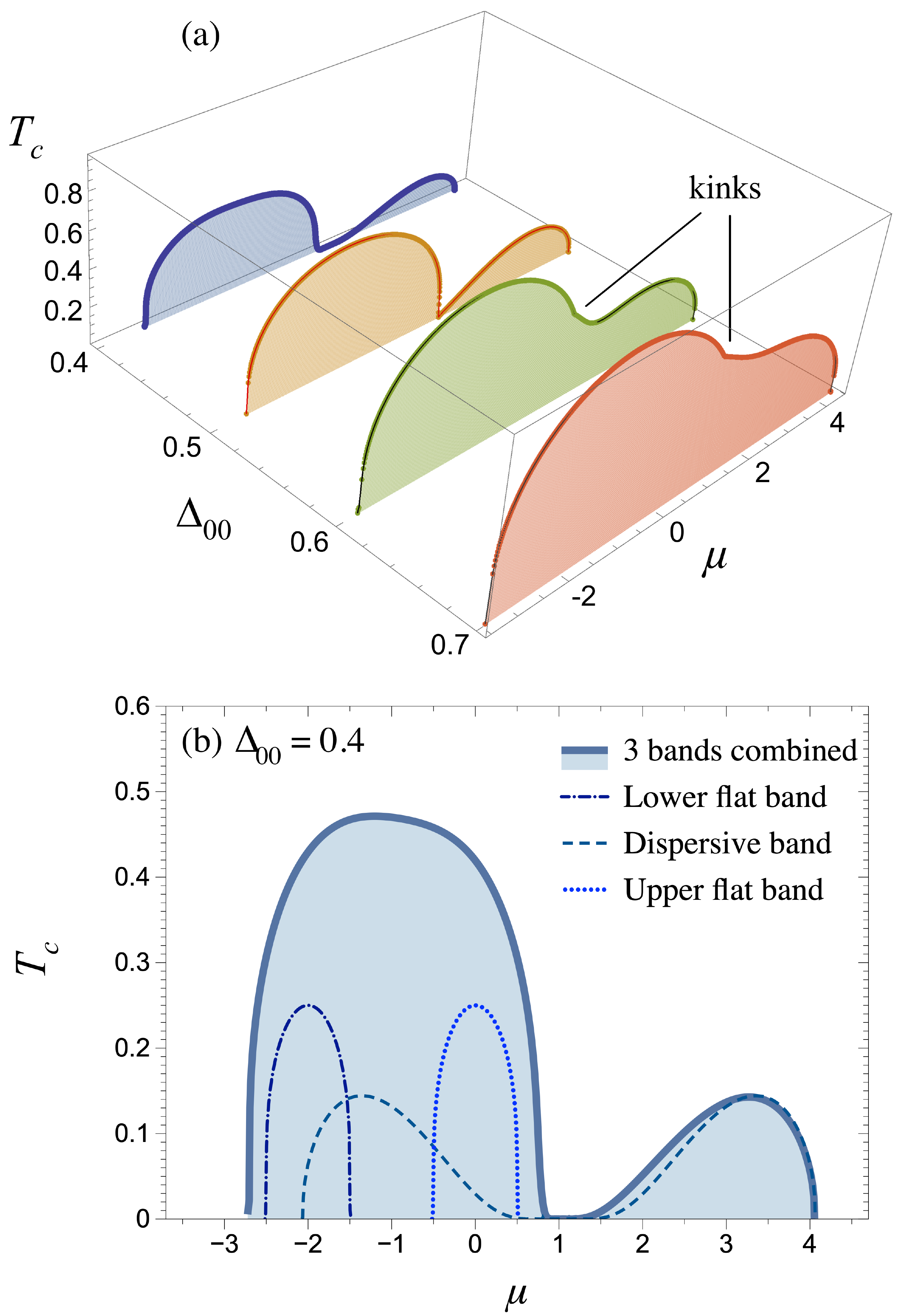}
  \caption{(Color online)
  a) Superconducting critical temperature $ T_c $
  as a function of the chemical potential $ \mu $
  for several values of $ \Delta_{00} $.
  b) $ T_c $ calculated individually
   for each band contribution
   compared with the combined results
   for $ \Delta_{ 0 0 } = 0.4 $.
  Energies are expressed in units of $ t $.
  }
  \label{Tc.Vs.mu}
\end{figure}

The numerical results of
$ T_c $ as a function of $ \mu $
for several values of $ \Delta_{00} $
are presented in Fig.~\ref{Tc.Vs.mu}.

A unique feature
for the square-octagon lattice
is the kink at $ \mu = 1 $,
as indicated in Fig.~\ref{Tc.Vs.mu}(a)
for $ \Delta_{ 0 0 } = 0.6 $ and $ 0.7 $.
This is due to the contribution
of the dispersive energy
in Eq.~(\ref{Eq.EdispSand}),
where a kink appears
for the hyperbolic tangent at $ \xi_{ \mu } = 0 $
in Eq.~(\ref{Eq.Ib}),
inducing a kink at $ \mu = 1 $
in the phase diagram.
As $ \Delta_{ 00 } $ decreases,
a reentrant superconducting phase
with two domes arises
for $ \Delta_{ 0 0 } = 0.4 $ and $ 0.5 $.

Interestingly,
two-dome superconducting phase diagrams
have been observed
in several unconventional superconductors,
as in cuprates, heavy-fermions,
pnictides, chalcogenides
and others~\cite{Das2016}.
Different theories
have been proposed
to explain the suppression
of the superconductivity
between the domes
in those compounds.
For the particular case of La-based cuprates,
superconductivity is suppressed
at $ x \sim 0.125 $~\cite{Buchner1994},
hence being called the 1/8 anomaly,
and there is some consensus
that this is due
to the static stabilisation
of stripes~\cite{Tranquada1995}.
For the Ce-based heavy fermion compounds,
on the other hand,
it has been suggested that the
domes are related
to different pairing mechanisms~\cite{Yuan2014},
although a conclusion
has not been reached yet~\cite{Zeng2016}.
For the YBCO compound,
moreover,
a richer scenario emerges
as an external magnetic field
is applied.
In the absence of an external field,
the domes are merged,
which is similar
to the data for the heavy-fermion CeCu$_2$Si$_2$
and also to our results in Fig.~\ref{Tc.Vs.mu}(a)
for $ \Delta_{ 0 0 } = 0.6 $ and $ 0.7 $.
As an external magnetic field is applied,
there are two superconducting domes,
which closely resembles the data for
La$_{2 - x } $ Ba$_{ x } $CuO$_4 $,
at $ x \sim 1/8 $
and our results in Fig.~\ref{Tc.Vs.mu}(a)
for $ \Delta_{ 0 0 } = 0.4 $ and $ 0.5 $.
For 50 T,
only the second dome survives.
Two possible scenarios have been proposed
to explain the experimental data for YBCO~\cite{Ranshaw2015}:
independent pairing mechanisms,
each responsible for a superconducting dome,
as for the case
of the Ce-based heavy fermion compounds;
or a single pairing mechanism,
where $ T_c $ is reduced
in the region between the domes.
Presently, we argue that
the two-dome phase diagram
observed in unconventional superconductors
can be related to
a single pairing mechanism,
specially when
it is taken into account
the multi-band aspect
of the microscopic models
that describe them.

Indeed, the emergence
of these two superconducting domes
in the square-octagon lattice
is understood
upon inspecting Fig.~\ref{Tc.Vs.mu}(b),
where
$ T_c $
is calculated individually
for each band in Eq.~(\ref{Eq.GapEquation})
for the particular case of $ \Delta_{ 00 } = 0.4 $.
(The numerical results
for different values of $ \Delta_{ 0 0 } $
in this range are qualitatively the same.)
Notice that the first dome
is essentially the combination
of the three contributions
in Eqs.~(\ref{Eq.Eflat0})-(\ref{Eq.EdispSand}),
which are summed in a nonlinear way,
as in Eq.~(\ref{Eq.Tc}),
while the second dome
is mainly provided by the dispersive band
in Eq.~(\ref{Eq.EdispSand}).
We see that
a single underlying pairing mechanism
is responsible for the appearance
of the two disconnected superconducting domes
in
the square-octagon lattice.
Hence, we argue that
the two-dome phase diagram
observed in
some unconventional superconductors
can be related to
a single pairing mechanism as well,
specially when
the multi-band aspect
of the microscopic models
that describe them
is taken into account.

Another remarkable feature
can be observed
in Fig.~\ref{Tc.Vs.mu}(b)
when we consider
only the contribution of the flat bands
for $ \mu < 1 $.
In this case,
two superconducting domes
are separated by an insulating phase,
since the contribution
from the dispersive band
is disregarded.
This resembles the phase diagram
experimentally observed
in twisted bilayer graphene~\cite{MacDonald2019}.
However,
upon the contribution
from the dispersive band
sandwiched between the two flat bands,
the superconducting domes become intertwined
in the square-octagon lattice.

\begin{figure}[t!]
  \centering
  \includegraphics[width=1\columnwidth]
  {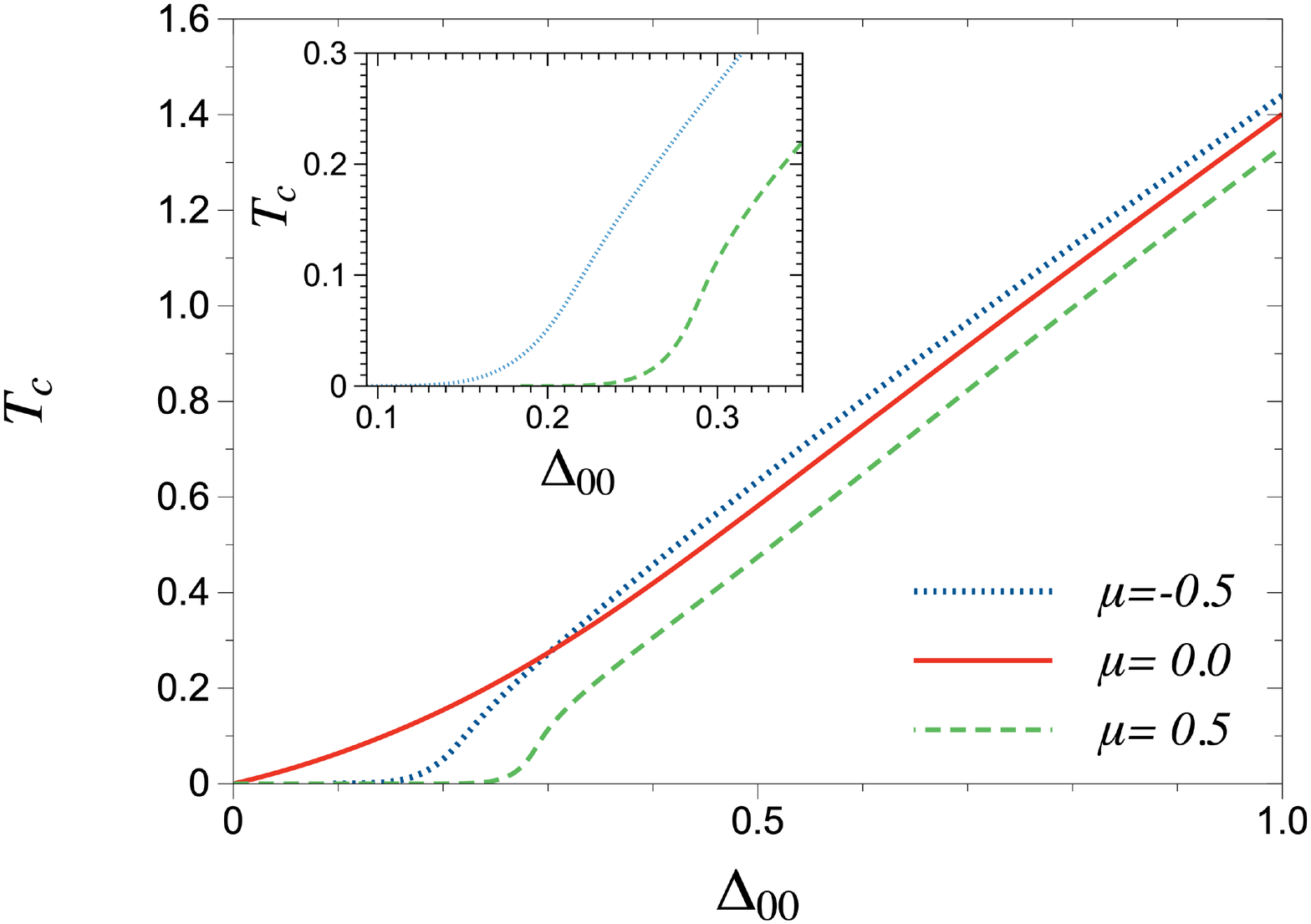}
  \caption{(Color online)
  Superconducting critical temperature $ T_c $
  as a function of the interaction strength
  $ \propto \Delta_{ 0 0 } $.
  Energies are expressed in units of $ t $.
  }
  \label{Tc.Vs.g}
\end{figure}

Next,
we concentrate
on the critical temperature $ T_c $.
The numerical results
for $ T_c $ as a function of $ \Delta_{00} $
for several values of $ \mu $
are presented in Fig.~\ref{Tc.Vs.g}.

There seems to be
an apparent threshold
in the interaction strength
for the appearance of superconductivity
at finite values of the chemical potential.
Per se,
this defines
a quantum critical point (QCP).
However,
the apparent suppression
of the critical temperature
for small values of $ \Delta_{ 00 } $
for $ \mu = -0.5 $ and $ 0.5 $
is simply a steep exponential decay,
as observed in the inset.
Indeed,
the absence of a QCP
at $ \mu = 0 $
in the square-octagon lattice
can be demonstrated as follows:
since the right-hand-side
of Eq.~(\ref{Eq.Tc})
is a positive, unbounded and monotonically decreasing
function of $ T_c $ at $ \mu = 0 $,
given any positive value for $ \Delta_{ 0 0 } $,
there is always a unique value for $ T_c $
that satisfies the self-consistent equation,
and, therefore
there is no QCP at $ \mu = 0 $.
Moreover,
while the contributions
related to the flat bands
are indeed bounded
for finite values of the chemical potential,
the contribution from the dispersive band
$ I( \mu, T_c  ) $ diverges
as $ T_c \rightarrow 0 $
(as demonstrated in the appendix)
and, hence,
there is no QCP
for the square-octagon lattice
even at $ \mu \neq 0 $.

Furthermore,
for larger values of $ \Delta_{ 00 } $,
there is a linear dependence
of $ T_c $ on $ \Delta_{ 00 } $.
Since $ \Delta_{00 } \propto g $,
the critical temperature
scales linearly
with the interaction strength.
Such linear dependence
was also obtained
for generic flat-band superconductivity~\cite{Heikkla2011,Kopnin2011}
%
and
should be contrasted with
the exponential dependence of $ T_c $
in conventional BCS superconductors,
$ T_c  \sim \exp \left[ - 1 / N(0) g \right] $,
where $ N(0) $ is the density of the states at the Fermi level.
From there on,
it has been conjectured that
the superconducting critical temperature
might reach room-temperatures
for a system presenting
flat bands~\cite{Volovik2018}.
Our results thus suggest
that high-$ T_c $
should be obtained in the square-octagon lattice,
specially in the vicinity of $ \mu = 0 $.

\subsection{Generic flat band system}

A superconducting QCP can be derived
for a generic flat band system as follows:
considering only the flat band
$ \epsilon = 0 $ in the Eq.~(\ref{Eq.Tc}),
the critical temperature satisfies
\begin{equation}
\tilde{T_c }
  =
  \frac{
 | \tilde{ \mu } |
  }{
  2 \tanh^{ - 1 } | \tilde{ \mu } |
  }
  \, ,
\label{Eq.Tc.FlatBand.B}
\end{equation}
where we have defined
$ \tilde{T_c } = T_c / \Delta_{ 0 0 } $
and
$ | \tilde{ \mu } | = | \mu | / \Delta_{ 0 0 } $.
The real values for $ T_c $
are constrained to $ | \tilde{ \mu } | < 1 $,
which is equivalent to
\begin{equation}
g > 2 | \mu |
\, ,
\end{equation}
assuming that
$ \Delta_{ 00 } = g / 2 $
from the Eq.~(\ref{Eq.Delta00}).
The above equation
defines a QCP whenever $ \mu \neq 0 $
and to the best of our knowledge,
this is the first time
that a QCP was derived
for flat-band superconductivity.
This is in contrast
with 2D Dirac fermion systems,
where there is a QCP at $ \mu = 0 $
and the threshold is absent
for $ \mu \neq 0 $~\cite{Marino2017}.
It is curious that precisely
the opposite applies
for the flat-band superconductivity.

Moreover,
assuming that
the only relevant contribution
in the self-consistent equation for $ T_c $
arises from the flat band
in Eq.~(\ref{Eq.Eflat0}),
and taking
$ \mu = 0 $,
we get that
$ \Delta_{ 0 0 }
$
satisfies the gap equation
and
$ \Delta_{ 00 } / T_c = 2 $,
which was obtained previously
for flat-band superconductivity~\cite{Heikkla2011,Kopnin2011}.
This ratio is comparable
to the results
obtained experimentally
for several high-$ T_c $ superconductors~\cite{Dagotto1994}.

\section{Conclusions}
\label{Sec.Conclusions}

In conclusion,
in this paper we investigated
the conditions for the appearance
of flat bands
in the square-octagon lattice.
We found
two perfectly flat bands
with a dispersive band between them
for equal values of
NN and NNN hopping
combined to
an external magnetic flux,
quantized as
$ \Theta = n \pi / 2 $,
for even values of $ n $.

Then, we introduced
an on-site attractive interaction
responsible for Cooper pairing
and calculated the
critical temperature
for the two-dimensional system.

The superconducting phase diagram
resembles those measured
for unconventional superconductors~\cite{Das2016}.
In the square-octagon lattice,
the disconnected superconducting domes
are provided
by the same underlying pairing mechanism
and are related
to the multi-band aspect of the system.
We argue that
the two-dome phase diagram
observed in unconventional superconductors
can be related to
a single pairing mechanism as well.

Despite the steep reduction of $ T_c $
for small values of the superconducting interaction strength,
there is no QCP
for the square-octagon lattice,
but an exponential decay of $ T_c $
for $ \mu \neq 0 $,
given the presence of a dispersive band
between the two flat bands.
In contrast,
we have derived
the QCP for flat-band superconductivity,
which occurs whenever $ \mu \neq 0 $.
To the best of our knowledge,
this is the first time that
such a QCP was derived.

The linear dependence
of the critical temperature
on the coupling,
as previously obtained
for flat-band superconductivity~\cite{Heikkla2011,Kopnin2011},
suggests that $ T_c $
might reach very high temperatures
in the square-octagon lattice.

The system discussed here
could be experimentally realised
using ultracold quantum gases in optical lattices.
Interactions can be easily implemented
using Feshbach resonances,
but it might be challenging
to design a laser setup
that yields a square octagon lattice,
and even more difficult,
NN and NNN hopping of equal intensity.
This problem can be easily solved
by using electronic quantum simulators~\cite{Slot2019}.
In this case,
lattices of any geometry
can be promptly realised on the nanoscale,
and the values of NN and NNN hopping
can be controlled independently,
with extreme precision.
This occurs because the overlap integrals
are designed by adjusting potential barriers,
instead of the distance between lattice sites.
Nevertheless,
the currently investigated electronic quantum simulators
rely on patterning adatoms
on the surface state of copper,
which is a non-interacting 2d electron gas,
and interactions are not yet tunable.
Further progress in the development of this platform
is required to observe the phenomenon discussed here
in designed structures.

Although we have focused
the discussion of superconductivity
on the square-octagon lattice,
some results are generic,
and valid for any system displaying flat bands.
Indeed, we generalised the calculations
in several instances
to models with a generic flat band,
and found that our results
confirm previously obtained ones in some limits,
and expands them.
In particular, we find a striking similarity
with the phase diagram observed experimentally
in unconventional superconductors.
Our studies might then
shed further light
on our understanding
of these complex materials.

\appendix

\section{Absence of a QCP}
\label{SecW}

Here,
we demonstrate
that the contribution
from the dispersive band
is divergent
when $ T_c \rightarrow 0 $
and therefore
there is no QCP
for the square-octagon lattice.
Indeed,
the integral
in Eq.~(\ref{Eq.Ib})
can be broken in two parts,
given the definition of the modulus.
Employing the changes of variables,
\begin{equation}
y_{ 1, 2 }
=
\pm
\left(
\frac{
 a_{ \mu } - 3 \sqrt{ 1 - x }
}{
2 T_c
}
\right)
\, ,
\label{Eq.y}
\end{equation}
where
$ a_{ \mu } = \left| 1 - \mu \right| $,
Eq.~(\ref{Eq.Ib}) can be rewritten as
\begin{eqnarray}
I( \mu, T_c )
& = &
\frac{ 2 }{ 9 }
\left[
\int_{ 0 }^{   a_{ \mu } / 2 T_c  }
dy_{ 1 }
\left(\,
\frac{ a_{ \mu } }{ y_{ 1 } }
-2 T_c
\,\right)
\tanh y_{ 1 }
\right.
\nonumber \\
& &
\hspace{-1.25cm}
+
\left.
\int_{  0 }^{ \left( 3 - a_{ \mu }  \right) / 2 T_c  }
dy_{ 2 }
\left(\,
\frac{ a_{ \mu } }{ y_{ 2 } }
+ 2 T_c
\,\right)
\tanh  y_{ 2 }
\right]
\, .
\end{eqnarray}
The r.h.s.
in the above expression
contains four terms,
the second and the fourth terms
are proportional to $ T_c $
and bounded in the limit
$ T_c \rightarrow 0 $;
the first and the third terms,
on the other hand,
are proportional to
\begin{equation}
\int_{ 0 }^{ \frac{ \alpha }{ T_c } }
dy \,
\frac{ \tanh y }{ y }
\, ,
\end{equation}
where $ a $ is a constant,
which diverges at
$ T_c \rightarrow 0 $.
It follows immediately that
$ I( \mu, \, T_c ) $
is divergent
and there is no QCP
for the square-octagon lattice,
due to the presence
of a dispersive band sandwiched
between the two flat bands.

\begin{acknowledgments}
L. H. C. M. N. thanks Prof. Cristiane Morais Smith and her research group for their kind hospitality during his stay at Utrecht University.
The authors thank Dr. Biplab Pal for helpful discussions,
and Rodrigo Arouca de Albuquerque
for a critical reading of the manuscript.
\end{acknowledgments}

\bibliographystyle{elsarticle-num.bst}
\bibliography{natbib}


%
%
%
%
%
%

\end{document}